# NOW THE DARK ELECTRON MULTIPLIER DOES SENSE DIRECTION OF THE DAEMON MOTION

E.M. DROBYSHEVSKI, M.E. DROBYSHEVSKI and V.A. PIKULIN

*Ioffe Physical-Technical Institute of RAS, 194021 St-Petersburg, Russia*
*E-mail: emdrob@mail.ioffe.ru*

Detection of the September maximum in the primary near-Earth daemon flux at high (~60°) Northern latitudes by our set-up with a plane horizontal scintillator is plagued by purely geometric factors; indeed, because of the Earth's rotation axis being tilted, the daemons catching up with the Earth in outer Near-Earth, Almost Circular Heliocentric Orbits (NEACHOs) strike the Earth along close-to-horizontal paths. Nevertheless, application of only two oppositely oriented, specially designed "dark electron multipliers" of the type TEU-167d (only their ø125-mm front disc is coated on the inside by a thick, ~0.5 μm Al layer, which permits such multipliers to detect primarily daemons flying inside them from the base to the disc) has made it possible for us to detect in one experiment, at a confidence level of $>3\sigma$, a flux of daemons captured from NEACHOs into Geocentric Earth-Surface-Crossing Orbits, as well as to record a decrease in the velocity of these objects from ~10 to ~7 km/s in a characteristic time of ~1 month resulting from their being slowed down in transits through the Earth's body.

## 1. Introduction. On the Interaction of the Daemon with Matter

Any object can be detected only through its interaction with matter. The fairly strong interaction with matter of daemons, these negative, multiply electrically charged objects of tentatively Planck nature [1-5], makes the task facing a researcher a far from trivial, many-faceted problem. We are going to consider (and use) here only some of its aspects.

Electromagnetic capture by a negative daemon, say, of a Zn nucleus in the ZnS(Ag) scintillator gives rise to excitation of the nucleus, with the ensuing ejection of a host of particles (electrons, nucleons, γ's) and generation of scintillations similar in shape to the ones caused by α-particles (Heavy Particle Scintillations - HPSs) [6]. Such capture of a nucleus in a fairly thick internal metallic (e.g., Al) coating of a PM tube also initiates emission of electrons. These electrons, similar to the photoelectrons, undergo in a PM tube trivial multiplication, which gives rise to formation of an electric signal at the instrument output [7].

In our previous publication [8], we have reported on a successful detection of daemons with a TEU-167 (Dark Electron Multiplier, DEM), a modified FEU-167 photoelectron multiplier, in which not only the rear (conical) and side (cylindrical) surfaces of the cathode section of the bulb, but the front disc (ø125 mm) as well, were coated on the inside with a fairly thick (~0.5 μm) aluminum film.

The results obtained in the experiment were very promising for the first attempt. These DEM experiments not only have corroborated the existence of a diurnal variation in the daemon flux from NEACHOs [8,9], which followed from celestial mechanics considerations, but offered also a lower estimate of its maximum morning level $f \geq 3.4 \times 10^{-7}$ cm$^{-2}$s$^{-1}$. The measurements were conducted in March 2009.

The response of the "symmetric" DEMs used by us was, however, practically independent of the direction in which the daemons crossed them. This was only to be expected - the DEMs respond primarily to objects that enter the Al coating from vacuum; in passing through vacuum where it cannot capture a new nucleus and become poisoned by it, a negative *c*-daemon (the daemon with the nucleus it has captured) is still capable of increasing substantially its negative charge in the course of successive daemon-stimulated nucleon decays in the captured nucleus [6-10], thus raising the probability (and energy) of capture of a new nucleus in the metallic layer. In our DEMs, the coating was applied, as prescribed by the Ref. [7] recommendations, to all of the internal surface of the cathode section of the bulb. Therefore, to resolve and amplify the effects associated with daemon transit through a DEM, we had to take into considerations additional factors which would make the detector properties asymmetric.

One of these factors was found to be the dependence of the amplitude-normalized area $S$ of an HPS signal excited in a thin, ~10 μm, ZnS(Ag) scintillator layer by a daemon that has captured a nucleus in it, on the direction in which it crosses this layer. We believe this to be a phenomenon unknown heretofore in experimental nuclear physics and originating only from the specific features of interaction with matter of a slowly moving (with $V \sim 10$ km/s) supermassive object which carries with it a nucleus excited in the process of capture. As the *c*-daemon emerges from the ZnS(Ag) layer into the substrate, a plate of polystyrene, far from all of the particles, both emitted by the nucleus excited in the capture and those created in secondary processes, can reach the scintillator. Some of the particles even become stuck in the polystyrene. Conversely, if the *c*-daemon together with the captured nucleus emerges from the ZnS(Ag) layer out into air or vacuum, nothing will stand in the way of the ejected particles to reaching the scintillator without losses. In the latter case, the scintillation will be slightly longer than in the preceding scenario. This dependence of $S$ on the direction of daemon propagation was pointed out by us as far back as 2003 [10].

Indeed, taking into account the correlation of $S$ with the direction of daemon motion through the detector, when the direction is determined by the sign of the time shift $\Delta t$ of an NLS (Noise-Like Signal) from DEM relative to the triggering HPS signal generated in the ZnS(Ag) layer turns out sufficient to estimate, say, the lower value of the excess in the number of events with $\Delta t > 0$ (the objects are falling downward) and $S < S_m$ ($S_m$ is the average HPS width) over the number of wide ($S > S_m$) events with $\Delta t < 0$, which can be identified with the upward daemon flux (provided the ZnS(Ag) layer is deposited on the upper surface of the polystyrene plate, and the plate itself is fixed at a distance of 29 cm over the DEM screen; for details, see below and [8]).

It thus becomes clear that detection by the same instrument of particles moving in opposite directions is not always desirable, for instance, at a high level of parasitic background signals. Indeed, in this case the presence in the $N(\Delta t)$ distribution the instrument is recording of several peaks (e.g., for $\Delta t > 0$ and $\Delta t < 0$) would lower the statistical significance of each of them.

Therefore, in the experiments being described below we made use of "asymmetric" TEU-167d which sense the direction of motion of the particles being detected. In these electron multiplier tubes, the fairly thick (~0.5 μm) internal Al coating was deposited on the inner surface of the front disc only [11]. The Al coating on the remaining surface of the near-cathode section was made as thin as possible (~0.05-0.1 μm); it provided electrical contacts and was only just opaque to light. The idea was to make a DEM be capable of reacting primarily to daemon propagation in the direction from the tube base to the screen. On the other hand, because the Al coating is evaporated on the inside surface of the glass, such DEM does not respond to the background produced by cosmic rays generating photons in the glass [12], so that with such multipliers in use underground experiments lose to some extent their potential.

## 2. Expected Specific Features of Different Daemon Fluxes in September

We conduct our experiments in the Northern hemisphere, at the latitude of $60^\circ$. In view of the fairly obvious, at least qualitatively, features of celestial mechanics evolution in the gravitational fields of the Sun, the Earth and other planets, and allowing for a certain loss of momentum incurred in the slowing-down transit through the Earth's body, daemons build up in NEACHOs exterior to the Earth's orbit [9]. Therefore, they catch up with the Earth in their motion, touching its orbit sometime ~10 days before the equinoxes [13].

The Earth's axis of rotation is tilted at an angle of 23.5$^o$ to the ecliptic plane. Therefore, in March the primary flux of NEACHO daemons with $V$ = 11.2-15 km/s falls on us primarily downward and reaches a maximum in evening hours (Fig. 1), a feature suggested already by the Baksan experiments as far back as 2006 and corroborated by our measurements performed with symmetric TEUs [8,9].

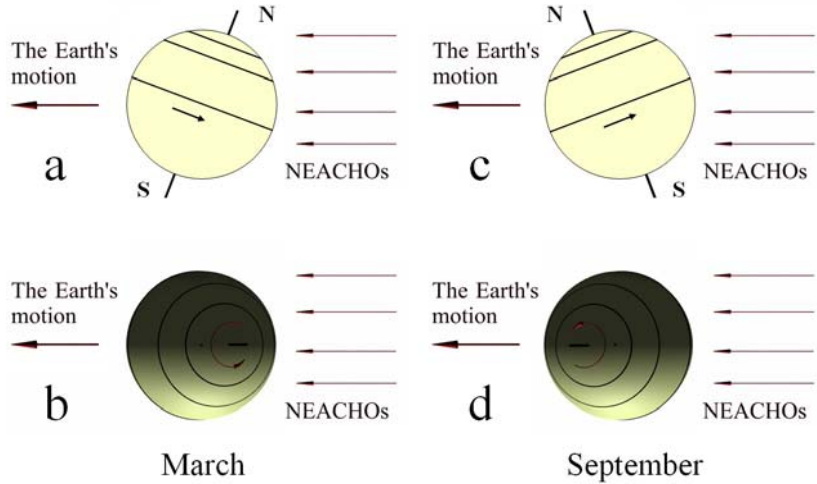

Fig. 1. Schematic of the fall on the Earth of the catching up NEACHO daemons during the spring (March) and autumn (September) maxima of their fluxes, as viewed from the Sun (*a* and *c*) and along the normal to the ecliptic plane from its north side (*b* and *d*). The equator, 40$^o$N and 60$^o$N latitudes are presented. The distortions of daemon trajectories near the Earth caused by its attraction are not shown; the trajectories should be curved toward the Earth, and the stronger, the lower is the daemon velocity relative to the Earth. We see nevertheless that in autumn the situation for daemon detection at high latitudes is not favorable, because a large part of daemons propagates here along the tangent to the Earth's surface.

In September, a similar situation is reproduced in the Southern hemisphere.

For us, in the Northern hemisphere, it does not look so favorable, which reflects the fairly obvious dependence of the NEACHO daemon flux on latitude. Although here the primary downward NEACHO flux likewise reaches a maximum in the evening hours, it does not cover all of the surface of the Northern hemisphere, with some objects, acted upon by the Earth's gravitation, propagating in trajectories close to tangents to the Earth's surface. This is why in September, operating with our detector having horizontal scintillation plates, we did not observe a clearly pronounced +30 μs maximum, which is seen to exist in the March $N(\Delta t)$ double events' distribution. The observed primary NEACHO flux is provided here by the daemons, which emerged upward with a velocity somewhat lower than that with which they struck the Earth (at lower latitudes) on the opposite evening side, and had already close to lost the memory of their original trajectory orientation. The resultant trajectories emerge from the ground primarily in the morning hours, an observation made in our Baksan measurements at the latitude of 43.2$^o$ [9]. It remains unclear what is their contribution at our latitudes.

This is why we focused our attention in data treatment on the night/morning hours (23$^h$ to 7$^h$), the more so that in this period electromagnetic interference was at its minimum (we have not thus far attempted any special optimization as to the time of day in data treatment).

The major contribution to formation of the ±30 μs maxima in $N(\Delta t)$ at our latitude (as we shall see in Sec. 4, with a shift of about a month) comes from the practically vertical (up- and downward) fluxes of daemons captured, as we believe, from the NEACHOs to Geocentric Earth-Surface-Crossing Orbits (GESCOs) as a result of their being slowed down by the Earth's material. We can discriminate here between two probable scenarios of capture, a slow and a fast one.

The slow scenario can be inferred from our observation in 2003 of a fairly long (~2 months) evolution of the GESCO population and comparison of this time scale with periods of close geocentric orbits of a few hours long. This suggested that the velocity of a daemon that had crossed the Earth with $V \sim 10$ km/s should have decreased by ~10 m/s only [10]. It thus followed that only the slowest of the NEACHO daemons, whose trajectories traverse the largest mass, i.e., close to the center of the Earth and along its diameter, can be captured into GESCOs. In the limit of the captured object passing through the Earth's center, its GESCO degenerates into a line coinciding with the Earth's diameter, which originally escapes beyond the Earth's surface. If the object had passed initially somewhat off the Earth's center, its GESCO will curve slightly and become open, because the object is moving now not in a point-mass gravitational field but rather through a body with the mass distributed over its volume [13]. It thus appears (now we are drawing from the results presented in Sec. 4 below) that the daemons captured initially from NEACHOs propagate along a fairly narrow bundle of GESCOs which do not reach high latitudes. This bundle has a waist close to the Earth's center and is originally oriented close to the Earth's orbit (the evening-morning direction). The GESCOs being not closed, their bundle spreads out gradually to drift eventually into high latitudes too.

The fast scenario draws to a considerable extent from the present experiments. It assumes that the daemon passing through the Earth loses immediately a few tens of percent of its kinetic energy relative to the Earth at infinity. As a result, far from all objects do return back on NEACHOs. A large fraction of them remains bound to GESCOs with apogees comparable with the dimensions of the Earth's gravitational sphere (and/or the distance to the Moon's orbit). Therefore, such daemons fall back on the Earth within a few days or weeks, with hardly any memory retained of the orientation of their original NEACHOs. Because of the high resistance to their motion through the Earth, a sizable fraction of the daemons will no more escape out of the Earth after their first backfall. And only some of them, that are crossing the outer layers of the Earth almost tangentially, may come out from it a few more times. This scenario also yields about a month or two for the characteristic lifetime of the GESCO population.

**3. September 2009 Experiment. Description of the Detector**

This time, we used only one modulus flushed through by nitrogen vapor. However, in addition to the standard polystyrene plate capped by a ~3.5 mg/cm$^2$ ZnS(Ag) layer of FS-4 scintillator powder and fixed inside a sheet-iron cube, 51 cm on a side, which was viewed from top by a standard triggering FEU-167, two DEMs of TEU-167d type were placed under the scintillation plate (Fig. 2).
These new modified DEMs described at the end of Sec. 1 [11], just as those employed by us earlier [8], were manufactured on our order by "Ekran - Optical Systems" Co (Novosibirsk). Their front discs positioned at base distance of 29 cm (standard for our experiments) from the ZnS(Ag) layer faced one another and were separated by a gap of ≈5 mm, in which a sheet of black paper was inserted (the latter was intended as a light shield, because at the center of the Al films on the DEM discs ø11-mm windows were left (Fig. 3), which were needed for preliminary DEM calibration with light; recall that all of the internal surface of the bulb's cathode section is coated by a standard photosensitive Sb-K-Na-Cs film).

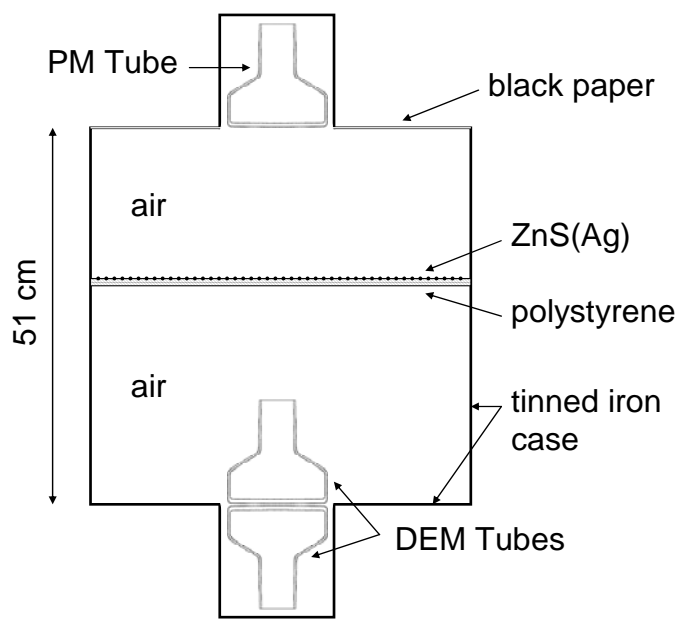

Fig. 2. Schematic of detector module. Two DEMs face one another with their front discs aluminum-coated on the inside, so that the bottom DEM (channel #22) senses upward moving daemons, while the top one (channel #24), those propagating downward. Fixed at 29 cm above their discs is the scintillation screen (ZnS(Ag) powder deposited on the upper surface of the polystyrene plate) which is viewed from a height of 22 cm by a standard FEU-167 producing a trigger pulse. The event is recorded if within a ±100-μs interval from the trigger HPS-pulse beginning there is a NLS from at least one of the DEMs.

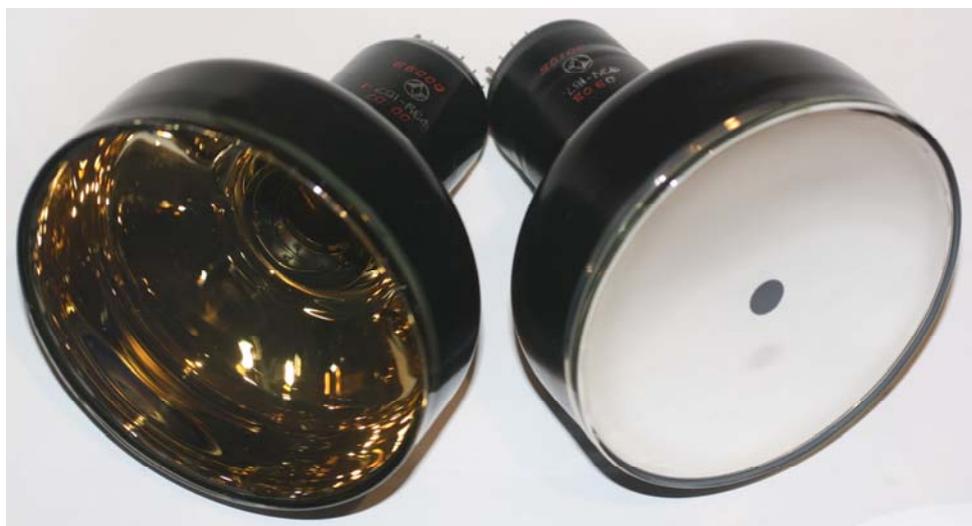

Fig. 3. PM tube of FEU-167-1 type (left; its front disc – photocathode – has ø125 mm), and DEM tube of new TEU-167d type [11] with its front disc only covered on the inside with thick ~0.5 μm Al layer; it has ø11 mm central transparent window for the preliminary light calibration of the instrument.

As always, the two S9-8 dual-beam digital oscilloscopes were triggered by the conventional PM tube [6-10], which, spaced by 22 cm from the scintillator, viewed it from above. The output signal of this PMT was fed to the first traces of the oscilloscopes (in subsequent data treatment, one considers only the HPS events; the corresponding triggering pulse was about 2.6 mV). Fed to the second traces of the two oscilloscopes are signals from the lower TEU-167d (No00086, its screen faces upward; it is channel #22) and the second

TEU-167d located inside the sheet iron box (No00105, its screen looks down; this is channel #24). The computer stores an event in its memory if signals appear on both traces of either oscilloscope. As in all our experiments, the setup recorded signals with a time shift within $\Delta t = \pm 100$ μs from the beginning of the triggering signal (the signals with $|\Delta t| \leq 0.4$ μs initiated primarily by cosmic ray showers in the atmosphere were vetoed).

We note immediately that channels #22 and #24 did not record any simultaneous events with $|\Delta t| > 0.4$ μs that would not look like cross-talk. This suggests that our DEMs do indeed respond primarily to "nonpoisoned" daemons that passed through their vacuum.

## 4. Main Results. The First Test of the Direction-Sensitive DEMs Produced Impressive Results

The observations were started practically immediately after our receiving the TEU-167d on August 31, 2009 from the Manufacturer, and were conducted from September 3 to November 2, 2009.

The most significant indication evidencing the existence of daemons is furnished by their flux from NEACHOs. The velocity of objects in this flux is confined within the interval of 10(11.2)-15 km/s. In our detector with a base dimension of ≈30 cm it produces in the $N(\Delta t)$ event distribution a maximum in the bin with a time shift $20 \leq |\Delta t| \leq 40$ μs (to be called in what follows the 30-μs maximum). As follows from the discussion in Sec. 2, the most favorable time for its observation in the Northern hemisphere is March. Nevertheless, despite the unfavorable conditions of observation of this maximum in autumn at the latitude of 60°N (see Sec. 2 and Fig. 1), we had practically no alternative to testing our DEMs of the new type. Rather than losing time in vain and waiting half a year more, we made up our mind to focus attention on revealing any sign of the existence of the 30-μs maximum. Our efforts were rewarded - we did observe it with our new asymmetric DEMs! It appeared, however, slightly later than the NEACHO maximum - now why?

Table 1 lists data pertaining to the evolution of the ±30 μs maxima. They are given for week-long intervals and nighttime/morning hours (from $23^h$ in the evening to $7^h$ in the morning, when the primary NEACHO flux should presumably emerge from under the ground at our latitudes [9], and the electromagnetic interference is at its lowest). The fairly conspicuous differences between the numbers of double events $\Sigma_{22}$ and $\Sigma_{24}$ for channels #22 and #24 and different weekly intervals should be assigned not only to purely statistical fluctuations and/or possible deviations in the oscilloscope triggering levels but also to fairly frequent malfunctions in the system initiated by strong uncontrolled cross-talk, a factor mentioned earlier [8], so that the real time of normal operation of the system can be noticeably less than one week. This naturally imposes limitations on evaluation of the lower limit of the daemon flux while not affecting in any way our conclusions, which thus far are largely of a qualitative nature.

In addition to the total numbers $\Sigma_{22}$ and $\Sigma_{24}$ of double events triggered by HPSs from the upper FEU-167 which viewed the ZnS(Ag) scintillator, one can find in the table numbers of events in the --30 μs and +30 μs bins in the $N(\Delta t)$ distribution which are arranged in accordance with the width $S$ of the HPSs. In this way we amplify the detector asymmetry, just as we did it in the preceding paper [8]. The first in the row are the numbers $n_w$ of "wide" events with $S > S_m$, and the second, separated by "/", the numbers $n_n$ of "narrow" events with $S < S_m$ (where $S_m$ reflects the average width of the HPSs in the given experiment).

To amplify the indications for the existence of upward/downward vertical fluxes, it appears natural to add the number of wide events in the -30 μs bin (the flux is upward; these numbers in Table 1 are given in bold) detected by channel #22 (the screen of its TEU-167d is looking up, so that it should detect, if our idea is correct, primarily the upward flux) to that of narrow events in the +30 μs bin recorded in channel #24 (the screen of its TEU-167d faces downward; these numbers are likewise given in Table 1 in bold). This sum $\chi$ should now be compared with one half (because we actually select one half of the events, more specifically, events with either $S > S_m$ or with $S < S_m$) of the total number of double events $\Sigma$ in both channels. For no effect present, we would obtain $\chi = \chi_m = 0.1\Sigma/2$, because the number of 20-µs-wide bins in a ±100 µs interval is 10. The confidence level (C.L. = $(\chi - \chi_m)\chi^{-1/2}$) of this 30-µs maximum expressed in terms of the rms error $\sigma$ is displayed in the last row of Table 1 for the two-week intervals.

Table 1. Weekly characteristics of the ±30-µs maxima in $N(\Delta t)$ distributions for the evening/morning ($23^h$-$7^h$) periods of operation of channels #22 (DEM screen looking upward) and #24 (DEM screen looking down, see Fig. 2). $\Sigma_{22}$ and $\Sigma_{24}$ - numbers of double events with HPS on the first oscilloscope trace of each channel ($\Sigma = \Sigma_{22} + \Sigma_{24}$). Events in the ±30-µs bins are divided into wide, $n_w$ ($S > S_m$, normalized area $S$ of the HPS signal is larger than its mean value $S_m$ for the autumn series of experiments), and narrow ones, $n_n$ ($S < S_m$). The width of the daemon-initiated HPS correlates with the direction of its motion through the ZnS(Ag) scintillator film deposited on the polystyrene substrate (see text). Shown in bold is the number of wide events $n_w$ in the -30-µs bin for channel #22 which detects upward-moving daemons (it is these daemons that should produce wide HPSs) and the number of narrow events $n_n$ in the +30-µs bin for channel #24 (it detects downward-moving daemons which produce narrow HPSs in the ZnS(Ag) layer).

The last row contains the C.L. for two-week periods. The weak 30-µs maximum detected in the first half of September probably reflects existence of the primary (but strongly inclined) NEACHO flux (note a similar feature in Table 2), and the well pronounced maximum in the beginning of October could be attributed to arrival to high latitudes of a vertical flux of GESCO objects and/or to their fall from far perigees.

| Dates, 2009 | | 3-10 Sept. | 10-17 Sept. | 17-24 Sept. | 24 Sept. - 1 Oct. | 1-8 Oct. | 8-15 Oct. | 15-22 Oct. | 22-29 Oct. | 29 Oct. -2 Nov. |
|---|---|---|---|---|---|---|---|---|---|---|
| 22 channel | $\Sigma_{22}$ | 166 | 448 | 114 | 133 | 82 | 167 | 129 | 205 | 258 |
| | $n_w/n_n$ for -30 µs | **9**/7 | **25**/23 | **4**/3 | **11**/9 | **5**/10 | **17**/10 | **4**/5 | **6**/9 | **13**/15 |
| | $n_w/n_n$ for +30 µs | 7/6 | 16/20 | 3/6 | 5/9 | 5/5 | 6/9 | 3/7 | 9/9 | 12/17 |
| 24 channel | $\Sigma_{24}$ | 191 | 323 | 85 | 110 | 109 | 139 | 87 | 184 | 178 |
| | $n_w/n_n$ for -30 µs | 13/7 | 18/19 | 4/3 | 7/3 | 3/5 | 5/5 | 3/7 | 10/8 | 4/8 |
| | $n_w/n_n$ for +30 µs | 6/**10** | 14/**21** | 5/**7** | 5/**3** | 2/**11** | 4/**13** | 5/**7** | 10/**8** | 10/**10** |
| | $\Sigma$ | 357 | 771 | 199 | 243 | 191 | 306 | 216 | 389 | 436 |
| | C.L. | +1.07$\sigma$ | | +0.58$\sigma$ | | +3.12$\sigma$ | | -0.85$\sigma$ | | +0.25$\sigma$ |

We readily see that during the first four weeks (all through September) the significance of the ±30-µs maximum hardly reaches 1$\sigma$. But in the beginning (the first two weeks) of October, the C.L. rises to 3.1$\sigma$, after which this maximum in $N(\Delta t)$ is even replaced by a weak minimum.

Viewed in the context of the "fast" scenario for the buildup of the GESCO population (see the end of Sec. 2 above), this time break can be readily accounted for by the return of daemons from the remote perigees of their orbits.

Considering now the "slow" scenario, we can assume that the flux of daemons captured into the GESCO bundle in the first half of September first crosses the Earth's surface at low latitudes and along the Earth's orbit, and only after a week or two its members lose memory of their original direction (recall that they cross the Earth along open rosette-like trajectories) to the extent where part of them drift to our latitudes. On coming here, their up (*or* downward) detected flux can be deduced from data of Table 1 for the period from October 1 to 15, 2009 (336 hours) to be as low as $f \approx [(5+17+11+13) - 0.1(191+306)/2]/(115 \times 336 \times 3600 \times 2) \approx 0.7 \times 10^{-7}$ cm$^{-2}$s$^{-1}$ (here 115 cm$^2$ is the area of the front disc of the DEM, the last figure 2 in parentheses reflects the flux in one direction only, say, upward; because GESCOs lie within a narrow bundle, we disregard here also the fact that the data listed in Table 1 relate to one third a day only). Significantly, the decrease in the velocity of these daemons should be small enough (say, from ~11 km/s down to ~7.5 km/s) to leave them in the ±30 μs bins, but at the same time let them escape one week later as a result of the further drop in the velocity.

To check the above assumption, we constructed Table 2. While being identical in structure to Table 1, it is intended for ±50 μs bins, i.e., for objects moving with vertical velocities of 7.5-5 km/s only.

The data of Table 2 provide a reasonable evidence for the assumption of the GESCO daemon flux drifting with time in velocity space within either scenario of their origin. Indeed, while before October 15 the significance of the ±50 μs peak was negligible, it is exactly starting with mid-October that the ±50 μs peaks exhibited noticeable significance (~1.54$\sigma$), paralleled by a drop in significance of the ±30 μs peaks from 3.1$\sigma$ down to a negative value (suggesting a minimum). Thus, the idea that the GESCO daemon population undergoes drift in the spatial, temporal, and velocity domains with a characteristic time of ~2 months, as this was suggested as far back as 2003 [10], finds additional support. The results obtained delineate a framework for development of celestial mechanics scenarios for the formation and evolution of the GESCO population with due regard for the resistance of the Earth's material.

Table 2. Same as in Table 1, but for the ±50-μs bins (vertical velocity of the objects $V$ = 10-7.5 km/s) in the $N(\Delta t)$ distributions. Note the appearance of a maximum in the second half of October, immediately after disappearance of the ±30-μs maximum (see Table 1).

| Dates, 2009 | | 3-10 Sept. | 10-17 Sept. | 17-24 Sept. | 24 Sept. -1 Oct. | 1-8 Oct. | 8-15 Oct. | 15-22 Oct. | 22-29 Oct. | 29 Oct. -2 Nov. |
|---|---|---|---|---|---|---|---|---|---|---|
| 22 channel | $\Sigma_{22}$ | 166 | 448 | 114 | 133 | 82 | 167 | 129 | 205 | 258 |
| | $n_w/n_n$ for -50 μs | **15**/5 | **26**/26 | **5**/6 | **10**/4 | **3**/5 | **12**/10 | **9**/6 | **13**/7 | **13**/10 |
| | $n_w/n_n$ for +50 μs | 8/4 | 23/18 | 6/10 | 6/8 | 5/5 | 6/8 | 8/4 | 10/5 | 12/9 |
| 24 channel | $\Sigma_{24}$ | 191 | 323 | 85 | 110 | 109 | 139 | 87 | 184 | 178 |
| | $n_w/n_n$ for -50 μs | 11/16 | 13/25 | 4/1 | 8/5 | 8/5 | 6/5 | 8/5 | 10/10 | 14/5 |
| | $n_w/n_n$ for +50 μs | 12/**9** | 8/**14** | 5/**2** | 3/**5** | 6/**6** | 5/**5** | 4/**4** | 5/**14** | 4/**9** |
| | $\Sigma$ | 357 | 771 | 199 | 243 | 191 | 306 | 216 | 389 | 436 |
| | C.L. | +0.95$\sigma$ | | -0.01$\sigma$ | | +0.03$\sigma$ | | +1.54$\sigma$ | | +0.04$\sigma$ |

## 5. Some Conclusions and Prospects

Application of DEMs which not just respond to daemons but sense their motion direction has broadened noticeably the potential of our detector. Indeed, even in September, a month unfavorable for the high latitudes of the Northern hemisphere, we have been able to reveal with one (strictly speaking, two) detector module the 30-μs maximum at a C.L. of 3.1$\sigma$, which exceeds markedly the C.L. = 2.85$\sigma$ reached in the March 2000 experiment with four modules and under considerably lower background, the best one-season result until now. Data were also obtained suggesting a velocity drift of the flux of daemons - presumably, of their GESCO population. We have grounds to hope that application of such direction-sensing DEM tubes in the more favorable conditions of the coming March will provide new intriguing information on the primary flux of the NEACHO objects.

We have to add, however, that already these first experiments suggest that the possibilities inherent in refining daemon-sensitive DEMs are far from being exhausted. A cursory examination of Table 1 reveals that the central DEM (channel #24) responds to the direction of daemon motion slightly better than the lower one (channel #22). It is quite possible that the rear Al coating thickness of the latter is not small enough, with the result that a negative $c$-daemon entering it from vacuum will be able to capture sometimes an Al nucleus here as well. One should therefore pay special attention to selecting a technology that would provide application of coatings more uniform in thickness and, just as obviously, to choosing materials (maybe, different ones?) for the front and rear coatings, as well as (also an important point) to their being free of radioactive contamination.

Accumulation of new results, paralleled by a continuous increase of their significance (now to >5$\sigma$), and a mutually noncontradictory refinement of previous scenarios and working hypotheses, including development on this basis of heretofore unknown instruments - all this testifies to the validity and basic strength of the original daemon paradigm. There are presently hardly any sound grounds for questioning it.

One could suggest a variety of ways for its further development. Some of them are pretty obvious - it is continuation of detection of daemons, near-Earth and (in the future) galactic ones, including upgrading the detectors themselves and development of celestial mechanics scenarios of evolution of the various daemon populations, in-depth understanding of the mechanisms and processes involved in their interaction with matter at different levels, starting with capture (and recapture) of nuclei to the catalysis of nucleon decay (or absorption?), on the one hand, and ending with formation of galaxies and of their active nuclei, on the other. Even attempts at grasping the consequences of application of the daemon paradigm, both to "grand" science, in cosmology and for development of Planck physics (see, e.g., [14] on numerous references to the ~$10^{19}$ GeV scale), and to "down-to-Earth" technologies, including designing compact and ecologically clean nuclear sources of energy, make one gasp and wonder.

**Acknowledgments**

The authors would like to use this opportunity to thank N.A. Krylova, the Director of the PMT production division of the "Ekran - Optical Systems" factory in Novosibirsk, for ensuring prompt production of the DEMs. Assistance of P.B. Simonov in detector assembly is gratefully acknowledged.

**References**

 1. M.A. Markov, Elementary particles of maximally large masses (quarks and maximons), *Sov. Phys. JETP* **24**, 584-592 (1967).
 2. K.P. Stanyukovich, On a possible kind of stable particles in Metagalaxy, *Dokl. USSR Acad. Scis.* **168**, 781-784 (1966).
 3. S. Hawking, Gravitationally collapsing objects of very low mass, *Mon. Not. Royal Astron. Soc.* **152**, 75-78 (1971).
 4. A.G. Polnarev and M.Yu. Khlopov, Primordial black holes and supermassive particles, *Usp. Fiz. Nauk* **145**, 369-401 (1985).
 5. E.M. Prodanov, Daemon decay and inflation, *Physics Lett. B* **681**, 214-219 (2009); arXiv:0910.1769v1.
 6. E.M. Drobyshevski, Detecting the Dark Electric Matter Objects (DAEMONs), *Astron. Astrophys. Trans.* **21**, 65-73 (2002)
 7. E.M. Drobyshevski, Observation of the March maximum in the daemon flux from near-Earth orbits in the year 2005: New efforts and new effects, *Astron. Astrophys. Trans.* **25**, 43-55 (2006); arXiv:astro-ph/0605314.
 8. E.M. Drobyshevski and M.E. Drobyshevski, Towards special daemon-sensitive electron multiplier: Positive outcome of March 2009 experiment, *Mod. Phys. Lett. A* (in press) (2010); arXiv:0912.4826.
 9. E.M. Drobyshevski and M.E. Drobyshevski, Study of the spring and autumn daemon-flux maxima at the Baksan Neutrino Observatory, *Astron. Astrophys. Trans.* **25**, 57-73 (2006); arXiv:astro–ph/0607046.
10. E.M. Drobyshevski, M.V. Beloborodyy, R.O. Kurakin, V.G. Latypov and K.A. Pelepelin, Detection of several daemon populations in Earth-crossing orbits, *Astron. Astrophys. Trans.* **22**, 19-32 (2003); arXiv:astro-ph/0108231.
11. E.M. Drobyshevski and M.E. Drobyshevski, Vacuum electron multiplier for registration of directional motion of nuclear-active particles, *Application for the RF Patent* No 2009148557, priority of 18.12.2009.
12. M.E. Drobyshevski, Peculiarities of reaction of photoelectron multiplier tubes onto the background cosmic rays, *Tech. Phys.* **80,** No4 (2010).
13. E.M. Drobyshevski and M.E. Drobyshevski, Daemons and DAMA: Their celestial-mechanics interrelations, *Astron. Astrophys. Trans.* **26**, 289-299 (2007); arXiv:0704.0982.
14. S. Weinberg, *Cosmology* (Oxford Univ. Press, 2008).